\documentstyle[twocolumn,prc,aps]{revtex}
\input epsf
\begin{document}

\twocolumn[\hsize\textwidth\columnwidth\hsize\csname @twocolumnfalse\endcsname
\title{$\Psi '/\Psi $ ratio in Nucleus-Nucleus Collisions : \\
a Measure for the Chiral Symmetry Restoration Temperature ?}
                                                            
\author{Heinz Sorge, Edward  Shuryak, Ismail  Zahed }
\address{
  Physics Department, State University of New York at Stony Brook,
 NY 11794-3800}
\date{May 15, 1997}
\maketitle    

\begin{abstract}               
We argue that a  decrease of the chiral scalar meson mass is  responsible for 
re-creation of $\Psi '$ from $J/\Psi $ in ultrarelativistic nucleus-nucleus 
collisions. This causes the charmonium yields to freeze out at temperatures  
close to the chiral symmetry restoration  temperature  $T_c$. As a result   
$\Psi '/\Psi $  may serve  as a thermometer for  $T_c$ itself. Results in a 
detailed reaction model support the conjecture. They show good agreement with 
recent data of NA38 and NA50 for $J/\Psi $ and $\Psi '$  production in S on U  
and Pb on Pb collisions.
\end{abstract}

\vspace{0.1in}
]

\begin{narrowtext}   
\newpage 
  
   Recent measurements by the NA50 collaboration at CERN
    show a so-called anomalous suppression of $J/\Psi $ yields
   in non-peripheral  Pb(158AGeV) on Pb collisions 
    \cite{na50g}.
   The results have been interpreted as a  hint for 
    Quark-gluon plasma (QGP) formation in these reactions 
   \cite{Blaizot}-\cite{Armesto}.
  The  QGP can  ``ionize'' a $J/\Psi $  analogously to the    photo-effect  
 \cite{photoeffect} or even screen it out of existence \cite{Matsui}. 
   Other explanations,  dissociation in coherent gluon fields 
   \cite{Loh} 
    and collisions with secondary hadrons (comovers)  
  \cite{Gavin}-\cite{Capella}
    have also been put forward. 
   Generally speaking,  the discussion is centered about the
   question whether the observed suppression of charmonia
    is an initial or a final-state effect. 
    The issue cannot be considered as settled yet. 
   The arguments expressed  e.g.\ in \cite{Blaizot}-\cite{Wong}
   that the different trends in  S+U versus
   Pb+Pb suggest the appearence of a strong energy density-dependent
    destruction mechanism for charmonium states 
    could  be  tested  by repeating the experiment at lower beam
   energies.
  Here we assume that the charmonium states are indeed destroyed
  very early \cite{footnote}.  
   
   Naively, the fate of the different
  charmonium states is  independent from each other and is governed
   by their  size and binding energy. 
  The location of the   $2s$ excitation $\Psi '$(3686) is barely below the         
   threshold to the $D\bar{D}$ continuum 
  while the $J/\Psi $   with a mass of 3097 MeV is  deeply bound.
  It is natural to assume that the  $\Psi '$ is more easily destroyed.
  Indeed, NA38 has found that the 
   $\Psi '/\Psi $  ratio goes down continuously in S on U collisions
   with increasing centrality. However, this ratio is leveling-off  at around 4\% 
  in  non-peripheral   Pb on Pb reactions 
  according to the preliminary   NA50 measurements.
   Approximately the same value is seen in the most central S on U
   collisions. 
  This value is  roughly a factor of 4 lower than in $pp$ and $pA$
   reactions. 
   
  Does the  constancy  of  the $\Psi '/\Psi $  ratio  result from an accidental
  fine-tuning of the parameters  which govern the survival rates of
  each species?  Here we answer otherwise. 
   $\Psi '/\Psi $  is leveling off, because 
   observed $\Psi '$ are   {\em re-created} from  $J/\Psi$.  If so,
    $\Psi '/\Psi $  will therefore no longer decrease in more
  violent collisions (at RHIC and LHC).  
 The  value of the  $\Psi '/\Psi $  ratio reflects on the temperature $T$ of
  the medium provided the equilibration mechanism is sufficiently  robust.
 In  equilibrium,   $\Psi '/\Psi $ changes with $T$   exponentially as
  $\exp \left(-(M'-M)/T\right) (M'/M)^{3/2}$. 
  In fact,  $\Psi '/\Psi $  is very sensitive to small changes of $T$,
    because the mass difference $M' - M$ is much larger than the relevant
   temperatures.
 The measured  $\Psi '/\Psi $  ratio then suggests a freeze-out temperature
   around 170 MeV.
   Such a temperature is close to the expected transition temperature
   $T_c$ for chiral symmetry restoration. 
  
   What may the nature of the interactions which equilibrate the
    $\Psi '/\Psi $  ratio  be?  
   In the vacuum  $\Psi '$ decays predominantly into a
   $J/\Psi $ and two pions in s-wave.
   The decay width  is tiny  (141 KeV), and therefore
    the inverse process in the pion gas is simply too slow.
   Based on the matrix element
   for the decay process \cite{BrCa75} we can calculate 
    charmonium (de-)excitation  probabilities in collisions with pions. However, 
   even those are  not large enough on the time scale of nuclear collisions.
  When  $\Psi '\rightarrow J/\Psi \pi \pi $ decay  was studied first
  in the seventies  Schwinger et al. advocated an interpretation   of these
  decays through the intermediary  of a scalar resonance \cite{Schwinger75}.
  The transition matrix element is suppressed in the vacuum
   due to the magnitude of the scalar resonance mass. 
   The presumably large transition rate between  $J/\Psi $  and  $\Psi '$ 
   in nuclear collisions may then be a consequence of a dropping
   scalar mass. 
  The  emergence of a low mass
    sigma meson which becomes degenerate with the pions at
  $T_c$  is  responsible for  rapid transitions
  between  $\Psi '$  and $J/\Psi $ around this temperature. 
  The transition rates become very small at slightly lower temperatures
  due to the sensitivity of the transition rates to the sigma 
  mass. This means in turn that the  $\Psi '/\Psi $  ratio 
 in heavy ion collisions may serve as a rather clean measure of  $T_c$ itself.
 
 In the remainder of this Letter we estimate 
  the transition rates between
   $J/\Psi$ and $\Psi '$ in   thermal matter close to $T_c$.
  Subsequently, we will assess the main corrections to the equilibrium limit.
  Finally, we implement the $T$ dependent transition rates into a simple model
   for  nucleus-nucleus  collisions
   and compare the calculated  $J/\Psi$ and $\Psi '$  yields to the
   preliminary NA38 and NA50 data.
    
   Let us  consider first 
   charmonium transitions in a heat bath 
   based on the linear sigma model.       
  At $T_c$ the quark condensate vanishes and chiral symmetry gets restored.
  Seen from the hadronic side, the sigma  meson mass  decreases with
 temperature $T$, because the sigma  state has to become degenerate with
  the pion. The pion mass is 
   expected not to change much with $T$.  Its smallness is
   protected by the spontaneous break-down of chiral symmetry.
  It may be more suprising that a sigma model describes the
   relevant physics if one approaches $T_c$ from above
    \cite{Hatsuda},\cite{Gocksch}.
   QCD matter at very large temperatures is supposedly composed of a weakly
  coupled gas of colored quarks and gluons. 
  However, at temperatures  $T_c < T < 2T_c$,
    strong nonperturbative phenomena
    in the  QGP may be present, since the 
   QCD coupling constant $g$ is of order one.
   The  scalar-pseudoscalar 
   modes are rather special as far as chiral symmetry is concerned.
   Condensation is about to occur in the $\sigma $ channel while the
   pions will emerge as the Goldstone bosons of the broken phase at
   slightly lower temperatures.
   Calculations based on the instanton model  support
   the idea that correlations due to strong attractive forces
    and perhaps  bound states   are formed in the QGP            
   at $T \sim T_c$  \cite{instant}.
\begin{figure}
\vskip -0.4in
\epsfxsize=3.8in
\centerline{\epsffile{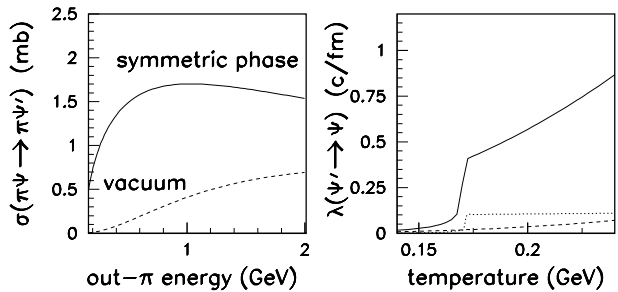}}
\vskip -0.05in
\caption[]{
 \label{figure1}
   $\sigma (\Psi \pi  \rightarrow \Psi ' \pi)$
    as a function  of the out-going pion  energy with the $\Psi '$
    at rest (left side). 
   The straight (dashed) line represents the calculation in the
    symmetric phase (vacuum with $m_s= m_s^0$).
    The right side shows the rate constant for
  $ \Psi '  $  transitions into a $ J/\Psi $  due to collisions and
  decays (straight line with dropping $s$ mass, dashed line with  $ m_s^0$).
   The dotted line represents just the contribution from  $ \Psi '  $ 
    decay in the chiral scenario.
}
\end{figure}

   Brown and Cahn  pointed out that the two 
   most important observations
   -- isotropic $\pi$ emission in their common rest frame and
    strong  enhancement at large invariant $\pi \pi$ masses --
   are reproduced by the linear sigma model at tree level \cite{BrCa75}.
 In this model  $\sigma $ and  $\vec{\pi} $   form a four-vector 
  $\vec{\sigma }=(\sigma ,\vec{\pi} )$ which enters into 
  the interaction terms of the Lagrange density 
 \begin{eqnarray}
     \label{lint}
   {\cal L}_{int}&= & c \Psi ' _\mu \Psi ^\mu  \: \vec{\sigma}^2
       - U(\vec{\sigma}^2)
   \quad .
 \end{eqnarray}
 A suitably chosen potential  $U$ 
 in eq.~(\ref{lint})
 gives rise to the spontaneous break-down
 of the symmetry  in which the isoscalar field
 acquires a nonvanishing  expectation value, in the  vacuum
  $\langle \sigma \rangle _0 $=$ f_\pi$.
  $f_\pi$$\approx$ 92 MeV is the pion decay constant.
  The effective potential  $U$  cannot be reliably calculated from QCD.
  The  standard  `Mexican hat' potential   was introduced
   by Gell-Mann and L\'evy \cite{linsm}   mainly for 
  its simplicity and renormalization property. 
  We are going to
   utilize the `freedom' in the choice of $U$
 to study  how a variation of $T_c$
  affects   $J/\Psi $  and $\Psi '$ in $AA$ collisions.
  In the broken phase
   the effective potential at finite $T$  is minimized by some  nonzero
   expectation value $f$ of the $\sigma $ field.
  With the field redefinitions $\sigma = (f+s) \cos \theta $
  and  $\vec{\pi} = (f+s) \hat{\pi} \sin \theta $
  (with $\hat{\pi}$ a vector of unit length)
  all burden of the self-interactions induced by $U$ is  laid
   on the   scalar-isoscalar field $s$.
   $s$ represents the fluctuations of the order parameter.
  The following interaction vertices 
   which are determined from the 
  Lagrange density enter into our tree-level  calculation:
 \begin{displaymath}
   V_{\Psi ' \Psi s} =  
          g M' \epsilon ' _\mu \epsilon ^\mu \quad ,
             \hspace{0.2em} 
   V_{\Psi ' \Psi \phi \phi} =  
          2 c \epsilon ' _\mu \epsilon ^\mu 
   \quad ,
 \end{displaymath}
 \begin{displaymath}
      V_{s\pi \pi}  =  -\frac{1}{f} \left(
                               2 p_1 p_2 + m_\pi ^2 \right)  ,
             \hspace{0.04em} 
      V_{ss\pi \pi}  =  -\frac{2}{f^2} 
                                p_1 p_2   .
 \end{displaymath}
  The Cartesian isospin indices of pions have been suppressed.
  The coupling constant  $g M' $ of $\Psi '$ decay equals $2 c f $.
  The field $\phi $ represents the $s$ meson field in the broken
   phase and an arbirary $\vec{\sigma}$ component 
  in the symmetric phase.
      
 The transition matrix element  
   ${\cal T}(\Psi ' J/\Psi \: \pi \pi) $
  depends at tree level  on $U$ only via the mass $m_s$ of the $s$ meson.
  We parametrize its  temperature dependence 
   by $(m_s(T)^2 -m_\pi ^2 )/(m_s^0 \:\mbox{}^2 -m_\pi ^2 )=
     \left(\langle  \bar{q}q \rangle (T)/\langle \bar{q}q 
                              \rangle _0 \right)^{2/3}$,
   because  $m_s$ is expected to track roughly
    the strength of the quark condensate $\langle  \bar{q}q \rangle$. 
   (Other parametrizations are also possible, without affecting the main results.)
   The $T$ dependence of the quark condensate  has been
    infered from   chiral perturbation theory ($\chi $PT)
    calculations  to order $T^6$ \cite{chpt}, 
    with the result that  the  condensate  vanishes 
    around 190 (150) MeV for the number of massless flavors $N_f$ 
   equal two (three).
    Several factors limit the applicability of $\chi $PT already below $T_c$.
   However, the $T_c$ values from this extrapolation are  overall consistent with
   the lattice QCD data  \cite{LATTICE}.    
  Later we will treat $N_f$  as  an  effective  number of
   degrees of freedom  whose magnitude    
    controls the value  of  $T_c$.
   Above $T_c$, the scalar  and the  pion mass are taken to be degenerate 
   and $T$-independent, since  only a 
   narrow window of temperatures above $T_c$ will be probed. 
    
  We fix the vacuum value of the $s$ meson mass to be 1400 MeV. 
   The particle data group lists a 
    very broad state at approximately this mass, the $f_0$(1400).    
   The location of the masss of the chiral sigma is controversial
   in the literature. Here we merely note that in the framework of
   the linear sigma model at tree-level a scalar meson with low mass 
   is strongly disfavored by experimental data on $\pi \pi $ scattering
    \cite{Gasser},\cite{Ecker}.
    This still allows a wide range of values for the scalar meson mass
    above 1 GeV. 
     (Note that the narrow scalar-isoscalar state just  below 1 GeV,
      the $f_0$(975),  is probably not the chiral partner of the pions but
      perhaps  a  $K \bar K$ ``molecule''.)
     Threshold theorems  are not well suited to constrain the
    $s$ mass much further, because the scalar propagator gets effectively
    replaced by a contact term. 
  We   determined   $g^2/4\pi\approx$0.36  
   from the partial decay width
   $\Gamma (\Psi ' \rightarrow \Psi \pi \pi)$.
    A term  $ i m^0_s \Gamma _s (m_{\pi\pi})$  containing the $s$ decay width
    has been added to the denominator of the
    free $s$ meson propagator  to take rescattering corrections into account.
  Similarly,  the $s$ acquires a nonzero width from   collisions
   ($s\pi \leftrightarrow s\pi$) at finite temperature. 
     
   On the left  hand side of Fig.~1 we compare the cross section  
   $\sigma (\Psi \pi  \rightarrow \Psi ' \pi)$
   in the vacuum and in the chiral phase.
    Note that the cross section is plotted as
    a function  of the out-going pion  energy. The in-going pion has   
   an energy larger by approximately 600 MeV which is  all of its
    typical  energy at relevant temperatures.
    It becomes apparent that the 
    thermally averaged cross section is tiny if the vacuum
     values are used. 
   On the right hand side the  rate constant
    $\lambda = 
       \Gamma _{dec}(\Psi ') + \Gamma _{coll} (\Psi ') $
    for  $ \Psi '  $  transitions into a $ J/\Psi $ 
     in the thermal heat bath 
    is displayed.  $T_c$=170 MeV has been 
   assumed. The total rate  shoots up at $T_c$ and increases smoothly
   afterwards.
   The most important contribution 
    to the transition between the two charmonium states 
    comes from the collisions with pions. 
   For comparison we also show the 
   contribution from the  $\Psi '$ decay to the rate  
    and the total rate but with
   the sigma  mass at its vacuum value.
\begin{figure}
\vskip -0.4in
\epsfxsize=3.3in
\centerline{\epsffile{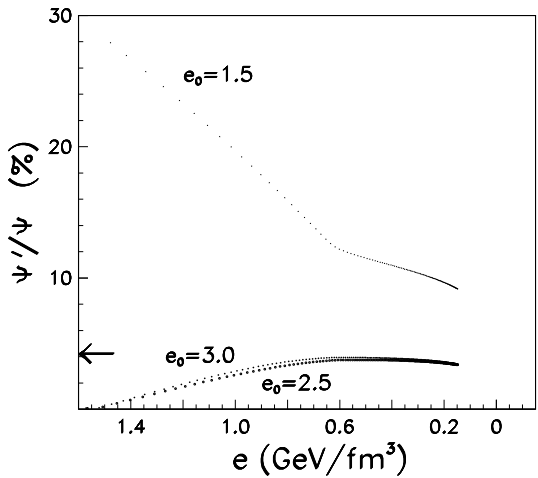}}
\vskip -0.05in
\caption[]{
 \label{figure2}
 Time evolution of  $\Psi '/\Psi $   for 3 different
 initial energy densities 1.5/2.5/3 GeV/fm$^3$
 and  $T_c$=170 MeV. 
  The equilibrium value 
   at $T$=$T_c$
  is  indicated by an arrow.
}
\end{figure}

  Next we would like to study
  the corrections to the $ \Psi ' / \Psi $ ratio, 
   mainly feed-down from $\chi $ states into $J/\Psi $ after the
   collision and  non-equlibrium effects. For that
   we consider  a simple model which, however,
   contains the essential physics. 
    Specifically, we calculate the initial charmonium production based on  the 
    Glauber model for  $AA$ collisions at a given  impact parameter $b$.
   We relate  $b$ to the experimentally measured transverse energy $E_t$
    by  employing  the results  in Ref.~\cite{Gavin}.
  The initial  production yields of  $J/\Psi$,
  $ \Psi ' $ and  $\chi $  in nucleon-nucleon collisions 
 are taken as 23, 6.5 and 84.8 respectively. 
   Here the  two  $\chi $ states 
  are lumped together with average branching ratio 0.21 into  $J/\Psi$.
   The total normalization  allows
   direct comparison with NA38 and NA50 data referred to as 
    $B \sigma (J/\Psi)/ \sigma (DY)$ in  \cite{na50l}.
    A common primordial suppression of all charmonia 
   depending on parameter $L$ (see \cite{Gerschel})
 is assumed   as   in  \cite{na50g},\cite{Blaizot}. 
    We calculate the final yields by integrating the coupled 
  rate equations for the local densities of the charmonium states, e.g.\  
   for the $\Psi '$
 \begin{equation}     
   \label{rateq}
   dN'/d \tau = - \lambda (\Psi ' \rightarrow \Psi) (N' - N'_{eq})
                  - \lambda _{c\bar{c}} (\Psi ' ) N'
      \quad .
 \end{equation}     
  Boost 
 invariance is assumed with formation time $\tau _0$=2fm/c. 
   As motivated in \cite{Blaizot}  the 
  rate constants $ \lambda _{c\bar{c}} $   for charmonium dissociation 
   are taken as zero below 
  and infinity above some critical energy density $e_{cr}$.
    The $\Psi '$ equilibrium density $  N'_{eq}$ 
  is related to the $J/\Psi $ density $N$ via 
   $ N'_{eq}$=$ N (M'/M)^{3/2}$$ \exp \left( -(M'-M)/T \right)$.
    Thus we need to model the temperature
    evolution of the created medium.
   The initial energy densities
    are taken as proportional to the density of participants
    in the transverse plane as in \cite{Blaizot}, more precisely
    $e= d_{AA} (E_t/N_p)/\tau_0 dN_P/d^2r$. 
   The choices $d_{SU}$=2.2 and  $d_{PbPb}$=4.0 lead to maximum local
    energy densities of 2.3 and 3.5 GeV/fm$^3$ for the most central
   S+U and Pb+Pb collisions.
   The equation of state which governs the isentropic expansion
    simulates a resonance gas with   energy density and temperature
   related by a power law $e= a b T^{b+1}$ and parameters $a,b$
   taken from \cite{bebie92}.
   The latter choices have been made with an eye on the space-time
    evolution according to RQMD calculations \cite{sorplb96} which connects
  our simple model with successful phenomenology of $AA$ collisions.
   Evolution is stopped at temperature 140 MeV although the precise
   value is unimportant due to the small transition rates
    below $T_c$ (cf.~Fig.~1).
  
  The $e_{cr}$ values which determine the  energy density
   above which a charmonium state immediately dissolves
   or may not be formed at all
  are the remaining parameters of the model. They 
 have been adjusted to reproduce the centrality
  dependence of $\Psi '$ and $J/\Psi $ yields in S+U and Pb+Pb collisions
   ($e_{cr}$=3.2/1.5/2.4 GeV/fm$^3$ for  $\Psi $/$\Psi '$/$\chi$). 
 Fig.~2 illustrates  the effect of $e_{cr}$($\Psi '$) by displaying
 the time evolution of  $\Psi '/\Psi $   for different
 initial conditions. At initial  energy density  just below
  the critical value of 1.5 GeV/fm$^3$ the 
  $\Psi '/\Psi $  ratio drops sharply until the system cools down to
  the critical temperature (here $T_c$=170 MeV). However, the initial
  $\Psi '$ yield is so far off equilibrium 
 that the $\Psi '  \rightarrow \Psi$ transitions do not 
  equilibrate the yield. 
 What  $e_{cr}$ effectively does is to
  define rather crudely a core  and a corona region. 
   The larger than  equilibrium values of  $\Psi '/\Psi $  in the corona 
  result in 
   increasing $\Psi '/\Psi $ values for more peripheral reactions.
  In contrast, starting with larger energy
  densities above  $e_{cr}$  and thus zero initial
 $\Psi '$ yield we see  that the reactions equilibrate the $\Psi '$
  yield. So,
   $\Psi '/\Psi $  works well  as a thermometer, 
  provided the initial energy densities in collisions
  are  clearly above 2 GeV/fm$^3$.   
  
 Fig.~3 contains a direct comparison of the calculated  $\Psi ' $  and
  $J/\Psi $  yields for S+U and Pb+Pb collisions with the NA38 and
  NA50 data.   $\Psi '/\Psi $   is displayed for three
   different choices of $T_c$ (150, 160 and 170 MeV).
   We find excellent  agreement with the data for $T_c$=170 MeV. 
    The $J/\Psi $ yield is only very mildly affected by the
    $T_c$ variations. The strong thermal
    $\Psi '$ suppression factor $N'_{eq}/N$ works against
    the $\Psi  \rightarrow \Psi '$ transitions to be effective
    as a destruction mechanism for  $J/\Psi $'s.
    On the other side, we provide an explanation why      
   scaling from $p$+$A$ works so amazingly well for 
   the $J/\Psi $  in S+U although
    some of the final feed-down from $\Psi '$ is gone. As noted above
   the $\Psi '$  feeds the $J/\Psi$ yield in the corona of the reaction. Thus
  $\Psi '$ suppression is partially offset by enhancement of direct $J/\Psi$'s!
\begin{figure}
\vskip -0.4in
\epsfxsize=3.3in
\centerline{\epsffile{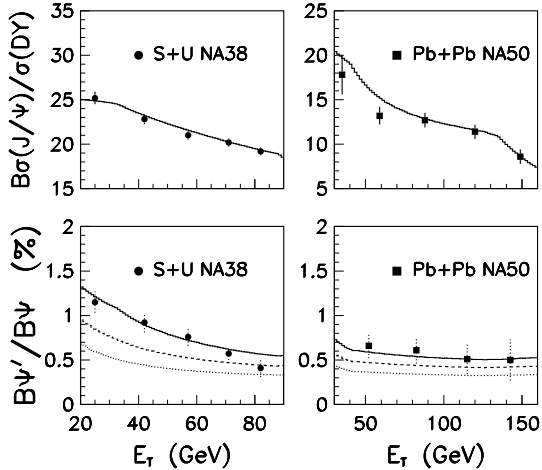}}
\vskip -0.05in
\caption[]{
 \label{figure3}
  Calculated $J/\Psi $ yield -- normalized to Drell-Yan cross section
   -- times branching ratio $B$ into $\mu^+\mu^-$
   versus preliminary data from  NA38 and NA50
   for S on U and Pb on Pb (upper panel),
   $\Psi '/\Psi $ ratio where each charmonium state is weighted
   with its decay probability into a dimuon pair
   versus data (lower  panel).  
  The calculated  $\Psi '/\Psi $ values are shown for three choices of
  $T_c$: 170 (straight line), 160 (dashed line) and 150 MeV
  (dotted line). 
}
\end{figure}
   Let us shortly comment on 
   feed-down from $\chi $ states. The feed-down corrections are sizable,
   even in central Pb on Pb collisions. The   $\Psi '/\Psi $  ratio 
   changes from 5.64 \% before feed-down to 4.07 \% after feed-down.
   We see that both surface emission and feed-down affect  $\Psi '/\Psi $ 
    by approximately 40 \%.  However, they contribute with different sign
   and cancel almost  each other. Furthermore, $T_c$ is rather well
    determined from  $\Psi '/\Psi $   even if one takes the feed-down
   corrections as a measure of the uncertainty involved. 
   To summarize, we  have suggested that 
   $\Psi '/\Psi $  levels off in ultrarelativistic nucleus-nucleus collisions, 
   because  $\Psi '$ is re-created from $J/\Psi $.
  Transitions between $\Psi '$ and $J/\Psi $ equilibrate the ratio  around $T_c$ 
  due to a  dropping  sigma meson  mass. 
   A chiral symmetry restoration temperature close to 170 MeV
   explains nicely why  $\Psi '/\Psi $ approaches a `universal' value
   of around 4 \%.
  The future dilepton experiments at RHIC and LHC will be important to test
  the suggested universality of the final  $\Psi '/\Psi $ ratio.
  
  
 H.S.\ thanks M.\ Gonin and C.Y.\ Wong for stimulating discussions.
 Both HS and IZ thank T. Hansson for an early discussion.      
This work has been
supported by DOE grant No. DE-FG02-88ER40388.

\end{narrowtext} 
  
\end{document}